# Machine Learning Extreme Acoustic Non-reciprocity in a Linear Waveguide with Multiple Nonlinear Asymmetric Gates

Anargyros Michaloliakos*[a], Chongan Wang[b], Alexander F. Vakakis[a]

[a] Department of Mechanical Science and Engineering,
University of Illinois, Urbana-Champaign

[a] Department of Mechanical and Aerospace Engineering, University of California, San Diego

*Corresponding author, am71@illinois.edu

## Abstract

This work is a study of acoustic non-reciprocity exhibited by a passive (i.e., with no active or semi-active feedback) one-dimensional (1D) linear waveguide incorporating two local strongly nonlinear, asymmetric gates. Strong coupling between the constituent oscillators of the linear waveguide is assumed, resulting in broadband capacity for wave transmission in its passband. Two local nonlinear gates break the symmetry and linearity of the waveguide, yielding strong global non-reciprocal acoustics, in the way that extremely different acoustical responses occur depending on the side of application of harmonic excitation, that is, for left-to-right (L-R) or right-to-left (R-L) wave propagation. To the authors' best knowledge that the present two-gated waveguide is capable of extremely high acoustic non-reciprocity, at a much higher level to what is reported by active or passive devices in the current literature; moreover, this extreme performance combines with acceptable levels of transmissibility in the desired (preferred) direction of wave propagation. Machine learning is utilized for predictive design of this gated waveguide in terms of the measures of transmissibility and non-reciprocity, with the aim of reducing the required computational time for high-dimensional parameter space analysis. The study sheds new light into the physics of these media and considers the advantages and limitations of using neural networks (NNs) to analyze this type of physical problems. In the predicted desirable parameter space for intense non-reciprocity, the maximum transmissibility reaches as much as 40%, and the transmitted energy from upstream (i.e., the part of the waveguide where the excitation is applied) to downstream (i.e., in the part of the waveguide after the two nonlinear gates) varies by up to nine orders of magnitude, depending on the direction of wave transmission. The machine learning tools along with the numerical methods of this work can inform predictive designs of practical non-reciprocal waveguides and acoustic metamaterials that incorporate local nonlinear gates. The current paper shows that combinations of nonlinear gates can lead to extremely high non-reciprocity while maintaining desired levels of transmissibility. This could lead to future investigation of how multiple nonlinear gates can be used as building blocks of designs incorporating robust passive acoustic non-reciprocity.

# 1. Introduction

Reciprocity is a basic feature of linear time invariant (LTI) elastodynamic systems such as mechanical oscillators and acoustic waveguides. To this end, the classical Maxwell-Betti reciprocity theorem states that the response of an LTI elastodynamic system is invariant when switching the points of excitation and measurement [1, 2]. Recently, there is growing interest in ways to break such reciprocity, given the important advantages that this would entail. Breaking reciprocity has been demonstrated utilizing both active and passive means, namely, by adding external biases [3, 4, 5], by incorporating time-variant properties [5], or, by imposing structural nonlinearity combined with some sort of configurational asymmetry. The latter option has the advantage of passivity over the prior two, as it does not require any sort of active or semi-active control nor any feedback energy source, and is the methodology adopted in this work. An added advantage of the nonlinear option for breaking non-reciprocity is the passive tunability of the resulting systems to energy and the wavenumber/frequency content of the applied excitations; this last feature is especially valuable when there is the need for the system response to be adaptive to non-stationary environmental inputs.

Focusing on acoustics, the break of reciprocity in an acoustic waveguide would violate the symmetry of the corresponding Green's function and would introduce interesting and potentially beneficial effects, the most obvious of which would be directional preference in wave transmission [6]. In turn, this would pave the way for various types of novel (and purely passive) acoustic elements such as acoustic diodes [7, 8, 9, 10], rectifiers [11], frequency convertors [12], topological insulators [13], and resonators with reconfigurable bandwidth properties [3]. In addition, there would be diverse applications in metamaterials incorporating acoustic logic and/or stress wave tailoring properties, vibration and shock isolation designs, and preferential energy redirection in complex systems, to name a few.

Local nonlinear elements integrated into otherwise LTI acoustic waveguides incorporating some form of configurational asymmetry may result in global non-reciprocal acoustics in a passive way, as it was recently shown in [15]. In that work the same 1D linear waveguide of coupled oscillators considered herein (but with weak intercoupling between oscillators) incorporating a single strongly nonlinear and asymmetric local gate was examined in the context of acoustic non-reciprocity. The nonlinear gate was composed of two dissimilar linear oscillators coupled through a cubic nonlinear stiffness, and machine learning was used to design the *local* gate for maximum *global* non-reciprocity. These results were extended to the more complex problem of the same gated waveguide but with strong intercoupling between oscillators in [17], where the machine learning study of the nonlinear acoustics was performed in a higher dimensional parameter space.

Based on the machine learning approach developed in [15, 17], we move forward to considering the more complex problem of linear acoustic waveguides incorporating multiple nonlinear gates and asymmetry (specifically, a series of two nonlinear gates), in order to show that the resulting non-reciprocity becomes much more intense compared to the case of a single gate examined in previous works. Hence, we aim to demonstrate that the introduction of multiple local nonlinear gates is a promising completely passive design tool as a means of introducing features of strong non-reciprocity in spatially extended linear acoustic waveguides. In addition, the current study allows us to extend the insight gained from the study of the waveguide with a single nonlinear gate, and assess the benefits gained in the nonlinear acoustics by introducing multiple such gates. However, the current problem is clearly more challenging and complex, since by introducing a cascade of nonlinear gates, the parameter space increases as well, thus requiring application of machine learning over broad frequency bands and higher dimensions. Moreover, there are specific features of the strongly nonlinear acoustics that complicate the untangling of the governing physics, e.g., the dependence of the responses on energy (or the applied excitation profile), and possible instabilities and bifurcations.

Nevertheless, as we will show in this work, the effort is well justified, as we report a logarithmic increase in the non-reciprocal effect compared to the case of the single gate, while maintaining similar levels of transmissibility, i.e., of energy transmitted downstream through the waveguide. The underlying reason behind this drastic enhancement in performance is the realization of targeted energy transfer (TET) between the two nonlinear gates, that is, local confinement (storage) of energy between the two gates, something that was not feasible before.

## 2. System Description and Parametric Analysis

A schematic of the 1D linear waveguide under investigation is shown in Figure 1. It is composed of two linear sub-waveguides (labelled L and R sub-waveguides) and incorporates a central region with two nonlinear gates and a linear coupling oscillator between them. Each sub-waveguide is composed of $N = 200$ identical grounded damped linear oscillators coupled to their neighbors through strong linear coupling stiffnesses, following the order shown in Fig. 1. Dissipation in the coupling stiffnesses is neglected for simplicity, since it can be considerably smaller than the grounding ones, as validated by the previous experimental research [14, 15]. The nonlinear gates consist of two linear oscillators coupled through an essential cubic nonlinear spring, with identical masses, damping coefficients but dissimilar grounding stiffnesses (and, hence, different linearized natural frequencies). In previous work [16], it is shown that this type of local asymmetry, together with nonlinearity, is sufficient to induce break of acoustic reciprocity in the response of the waveguide; that is, one would expect different responses when a single harmonic excitation is applied by switching the positions of the excitation and measurement. In Fig. 1 the harmonic excitation $F(t) = F_p \cos \omega t$ is applied to oscillator $p = 4$ of the L sub-waveguide, but in this study the excitation will be switched to an analogous position at the R sub-waveguide. Throughout this work we define the sub-waveguide that includes the oscillator where the excitation is imposed as the *upstream* one, and the other sub-waveguide as the *downstream* one.

The system parameters are denoted in Fig. 1, and are considered to be O(1) quantities. This implies that both sub-waveguides are composed of strongly coupled grounded oscillators, and the single passband of the corresponding linear waveguides of infinite extent (i.e., boundary-less) is broad [17]. Asymmetry in the waveguide is introduced by setting dissimilar values for the coefficients of the nonlinear stiffnesses of the two gates, namely, $k_{c1} \neq k_{c2}$, while set equal to $k_1$ the grounding stiffnesses of the 0-oscillators of the L and R sub-waveguides, (but note that they are detuned compared to the grounding stiffnesses of the other oscillators which are set equal to $k$, cf. Fig. 1).

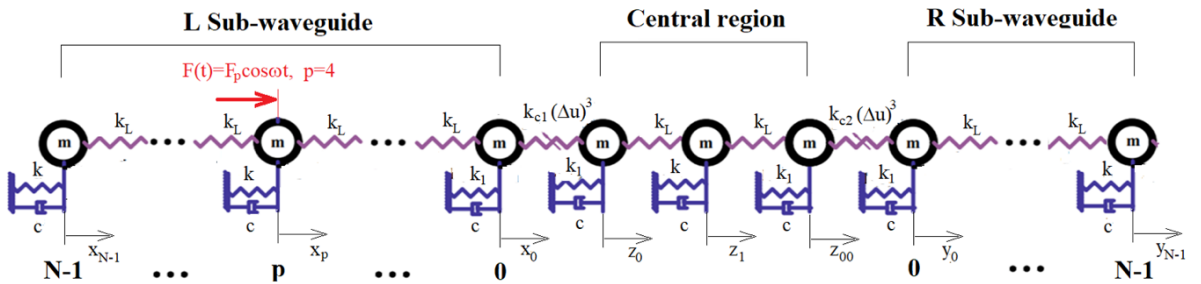

Figure 1. Schematic of the strongly coupled waveguide incorporating two nonlinear gates.

Considering the governing equations of motion, we denote by $\omega_0 = \sqrt{k/m}$ the natural frequencies of the oscillators in the linear sub-waveguides, and introduce the non-dimensional time $\tau = \omega_0 t$. Moreover, introducing the normalizations, $d = k_L/k$ (normalized intercoupling stiffness), $\xi = c/2\sqrt{mk}$ (critical viscous damping ratio), $\hat{\omega} = \omega/\omega_0$ (normalized excitation frequency), $A_p = F_p/(dk)$ (normalized excitation magnitude), $\alpha_1 = k_{c1}/k_L$ and $\alpha_2 = k_{c2}/k_L$ (normalized nonlinear coefficients), and $\sigma = (k_1 - k)/k_L$ (normalized stiffness detuning parameter), the equations of motion are expressed in the following normalized form with all initial conditions assumed to be zero,

$$x_i'' + 2\xi x_i' + x_i + d(x_i - x_{i-1}) + d(x_i - x_{i+1}) = [dA_p e^{j\hat{\omega}\tau} + cc]\delta(i-p), \quad i = 1,2,\ldots, N-1$$
$$x_0'' + 2\xi x_0' + \alpha(1+d\sigma)x_0 + d\alpha_1(x_0 - z_0)^3 + d(x_0 - x_1) = 0$$
$$z_0'' + 2\xi z_0' + (1+d\sigma)z_0 + d\alpha_1(z_0 - x_0)^3 + d(z_0 - z_1) = 0$$
$$z_1'' + 2\xi z_1' + z_1 + d(z_1 - z_0) + d(z_1 - z_{00}) = 0$$
$$z_{00}'' + 2\xi z_{00}' + (1+d\sigma)z_{00} + d\alpha_2(z_{00} - y_0)^3 + d(z_{00} - z_1) = 0$$
$$y_0'' + 2\xi y_0' + (1+d\sigma)y_0 + d\alpha_2(y_0 - z_{00})^3 + d(y_0 - y_1) = 0$$
$$y_n'' + y_n + 2\xi y_n' + d(y_n - y_{n-1}) + d(y_n - y_{n+1}) = 0, \quad n = 1,2,\ldots,N-1$$

(1)

where $j = (-1)^{1/2}$, $cc$ denotes complex conjugate, all displacements are functions of the normalized time $\tau$, and prime denotes differentiation with respect to $\tau$. Lastly, for convenience we introduce the normalized critical damping ratio $\zeta = \xi/d$.

In the forcing configuration shown in Fig. 1, the displacements of the oscillators of the upstream and downstream linear sub-waveguides are denoted by $x_p$, and $y_p, p = 0,\ldots,N$, respectively, whereas the oscillators of the central region by $z_0$, $z_1$ and $z_{00}$. The two nonlinear gates are formed between the response pairs $(x_0, z_0)$ and $(z_{00}, y_0)$, respectively. With the dissipation and excitation terms ignored, the normalized dispersion relation of the *infinite linear homogeneous waveguide* (without incorporating the two gates) is given by,

$$\Omega/\omega_0 = \sqrt{1 + 4d\sin^2(\theta/2)}, \quad 0 \le \theta \le \pi \tag{2}$$

where $\Omega$ and $\theta$ are the frequency and wavenumber of a propagating wave, respectively, and the limiting values of the frequency corresponding to $\theta = 0, \pi$ define the normalized passband of each sub-waveguide of infinite extent; this is the frequency range within which harmonic waves may transmit uninhibited through the acoustic medium. It follows that by selecting the normalized excitation frequency $\hat{\omega}$ within the normalized passband, the semi-infinite sub-waveguides can transmit harmonic waves in the broad band frequency range defined by $1 \le \hat{\omega} \le \sqrt{1+4d}$. As proved in [17], the nonlinear acoustics of the gated waveguide is then governed by the scattering – transmission and reflection – of these harmonic waveguides at the nonlinear gates. Lastly, note that $\hat{\omega}$ is expressed in terms of the wavenumber by the dispersion relation (2).

At this point, we proceed to investigate the nonlinear acoustics through numerical simulations. To this end, the normalized equations of motion (1) we numerically integrated for the system parameters provided in Table 1, and total simulation time $0 \le \tau \le 1500$. To highlight the different types of nonlinear acoustics realized in the waveguide three different configurations are considered (depending on system and excitation parameters), labeled as Systems 1,2 and 3. Since the nonlinear stiffnesses of the two gates provide the only source of configurational asymmetry, it holds that by exchanging the excitation position from the L to the R sub-waveguide is equivalent to interchanging the nonlinear coefficients $\alpha_1$ and $\alpha_2$; in that case the definitions of upstream and downstream sub-waveguides is also interchanged.

Table 1. The three waveguide configurations considered in the numerical simulations.

| Parameters | $d$ | $\{a_1, a_2\}$ | $\zeta$ | $\theta$ | $\sigma$ | $A_p$ | $p$ |
|---|---|---|---|---|---|---|---|
| System 1 | 0.5 | {0.15,0.3} | 0.013 | $(1/6)\pi$ | -1.5 | 1.0 | 4 |
| System 2 | 0.35 | {1.81,3.45} | 0.023 | $(2.5/6)\pi$ | -1.4 | 0.46 | 4 |
| System 3 | 0.4 | {3.9,3.1} | 0.013 | $(3/6)\pi$ | -1.4 | 0.4 | 4 |

In Figure 2, we depict the responses of System 1. The results for $(\alpha_1, \alpha_2) = (0.15, 0.3)$ (Left-to-Right wave transmission – L-R) are shown in Figs. 2a-d while the results when the excitation is switched between sub-waveguides, i.e., for $(\alpha_1, \alpha_2) = (0.3, 0.15)$ (R-L wave transmission), are depicted in Figs. 2e-h. In Figs. 2a,b and 2e,f, we compare the steady state displacement time series of the oscillators of the two gates $(x_0, z_0)$ and $(y_0, z_{00})$. For L-R wave transmission the steady state oscillation amplitudes of the gate oscillators before the first gate and after the second are similar, so it is inferred that the wave energy propagates through nonlinear central region consisting of the two gates. However, this is not the case for R-L wave transmission since drastically diminished energy transmits through the two gates; this is inferred from the fact that the amplitude of $y_0$ is much smaller compared to $x_0$ at steady state. Therefore, strong non-reciprocity is realized in the steady state acoustics, yielding nearly uni-directional energy transmission (specifically, R-L but not L-R). To explore further the nonlinear non-reciprocal acoustics, we proceed to postprocess the computed time series.

The harmonic content of the transmitted waves is studied by computing the amplitude of the discrete Fourier transform of the response $y_0$, the oscillator after the second gate for L-R and R-L wave transmission. In Figs. 2c and 2g, only a single dominant harmonic is noted at the frequency of the exerted harmonic excitation, which implies that in both directions the transmitted waves are (nearly) monochromatic at steady state. Furthermore, the corresponding energy measures presented in Figs. 2d and 2h provide a clear illustration of the non-reciprocal feature of the nonlinear acoustics. These measures are evaluated by computing the total work exerted by the applied external force, as well as the energy transmitted to the downstream sub-waveguide; this enables the study of non-reciprocity and transmissibility in the waveguide from an energy context. Following [17], the exerted work (input energy by the force) up to time $\tau$, $E_{input}(\tau)$, and the transmitted energy through the gate to the downstream waveguide up to time $\tau$, $E_{downL-R}(\tau)$ for L-R propagation and $E_{downR-L}(\tau)$ for R-L propagation, are given by:

$$\begin{aligned} E_{input}(\tau) &= \begin{cases} \int_0^\tau 2dA_p \cos(\widehat{\omega}\tau_0)\, x'_p \, d\tau_0 & (L-R) \\ \int_0^T 2dA_p \cos(\widehat{\omega}\tau_0)\, y'_p \, d\tau_0 & (R-L) \end{cases} \\ E_{downL-R}(\tau) &= \int_0^\tau d(y_0 - y_1) y'_0 \, d\tau_0 \\ E_{downR-L}(\tau) &= \int_0^\tau d(x_0 - x_1) x'_0 \, d\tau_0 \end{aligned} \qquad (3)$$

From Fig. 2d, we note that for L-R wave propagation the transmitted energy in the downstream R sub-waveguide is about 20% of the input energy, whereas it is almost zero for R-L wave transmission. These results showcase the strong acoustic non-reciprocity realized in System 1, allowing transmission energy unidirectionally (i.e., only L-R) through the gated waveguide. The nonlinear mechanism inhibiting energy transmission in the other direction (R-L) will be discussed later. Moreover, the acoustic non-reciprocity in this case is realized through nearly monochromatic waves at the steady state, and these can be analytically studied by the Complexification-Averaging methodology discussed in [17]; this, however, will not pursued any further in this work.

The corresponding numerical results for System 2 are shown in Figure 3. L-R wave propagation is considered in Figs. 3a-d, while R-L wave propagation in Figs. 3e-h. Again, energy transmission is only allowed in the L-R direction, revealing strong non-reciprocity in the acoustics. A key difference compared to System 1 is that for L-R propagation the responses have strong modulations and are referred to as strongly modulated responses (SMRs). This type of responses was also encountered in [17] and result due to the strong nonlinearities at the gates. This is signified by the side bands about the center excitation frequency at the Fourier spectrum of Fig. 3e, a clear indication of quasi-periodicity (modulation). On the contrary, for R-L propagation the responses are nearly monochromatic, resembling the responses of System 1.

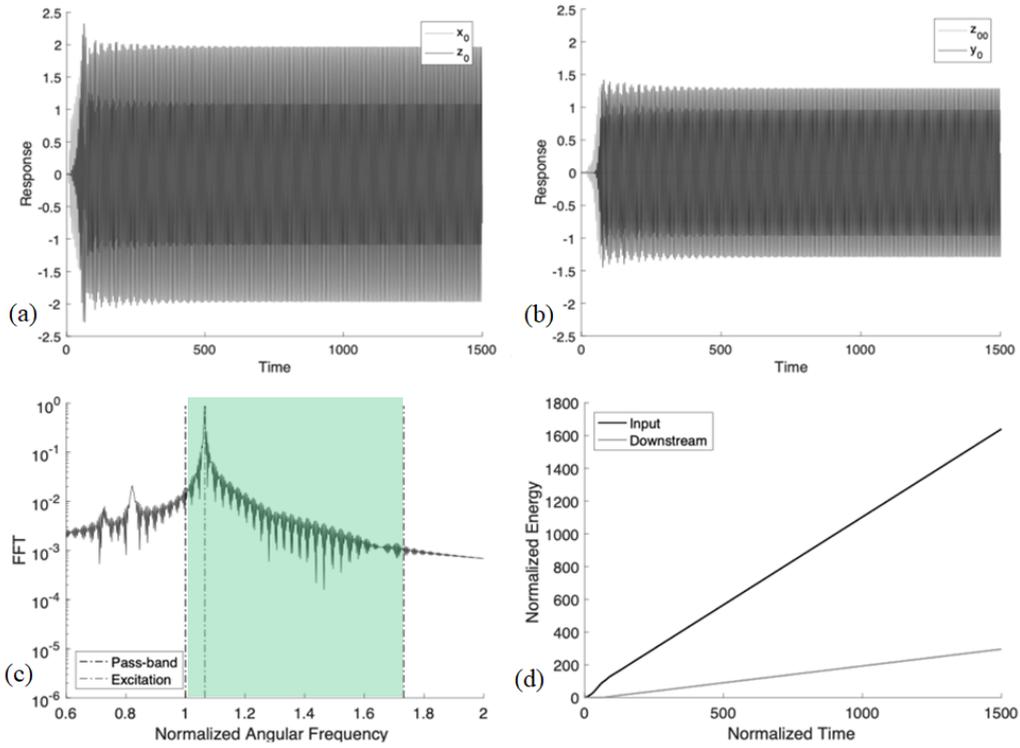
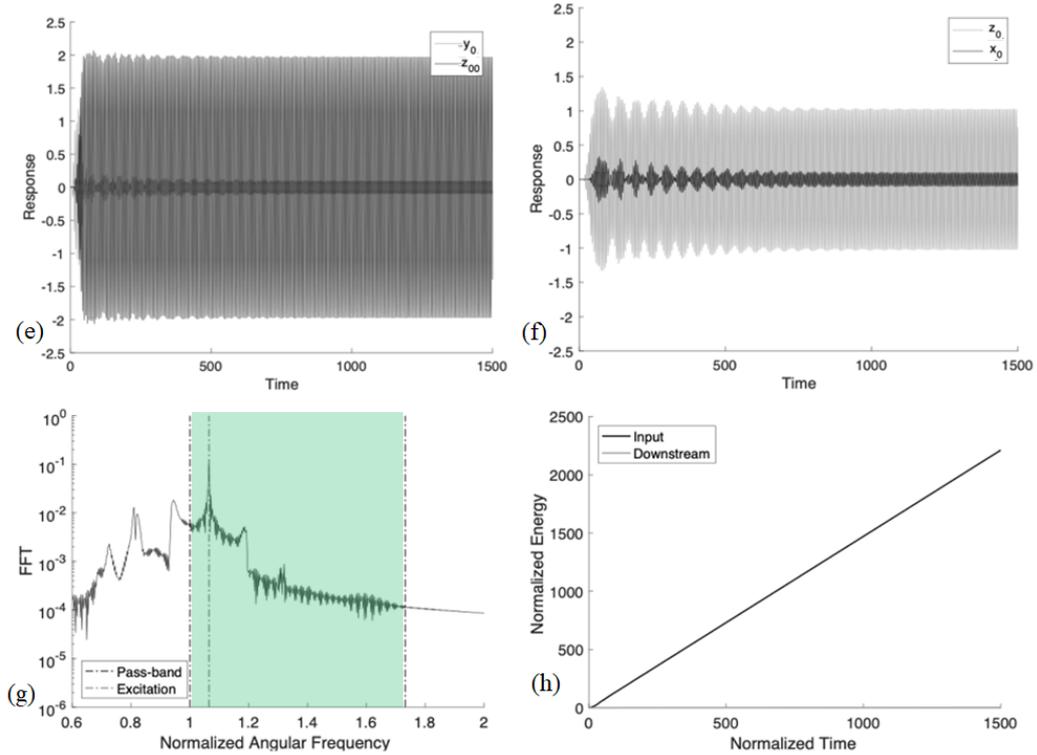

Figure 2. L-R wave propagation for System 1: (a,b) Time series of the oscillators of the nonlinear gates ($x_0,z_0$) and ($z_{00},y_0$), respectively, (c) Fourier spectrum of the oscillator $y_0$, and (d) energy measures; R-L wave transmission for System 1: (e,f) Time series of the oscillators of the nonlinear gates ($y_0,z_{00}$) and ($z_0,x_0$), respectively, (g) Fourier spectrum of the oscillator $y_0$, and (h) energy measures – the pass band and excitation frequency are indicated in Figs. (c) and (g).

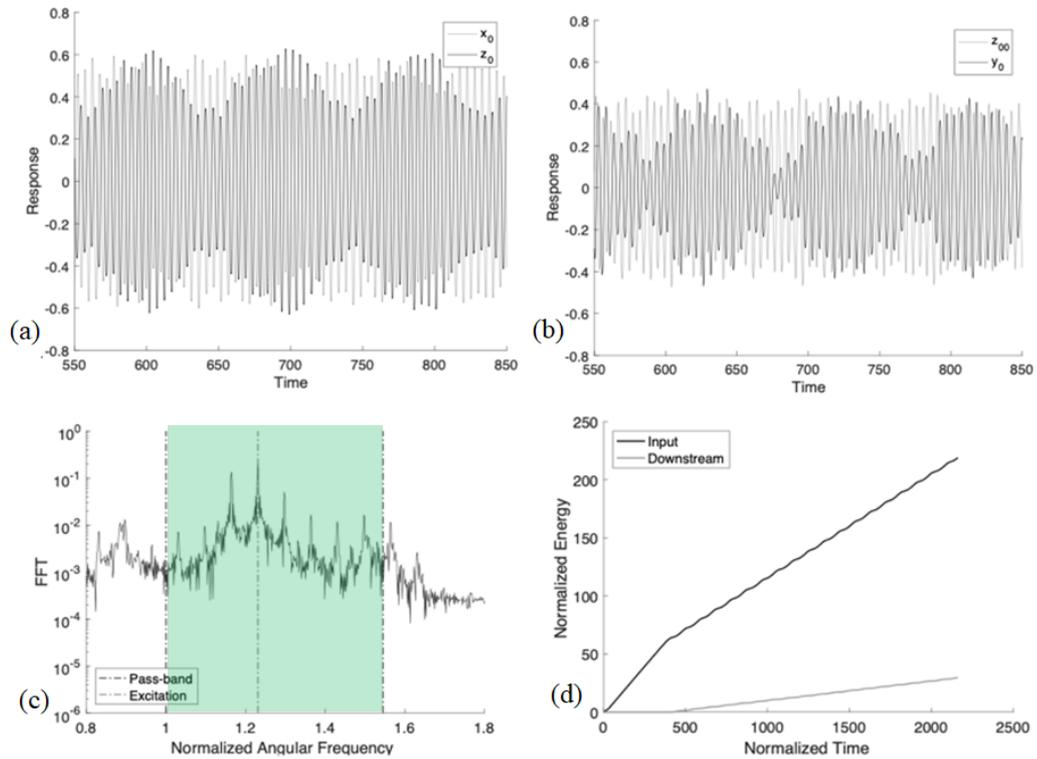
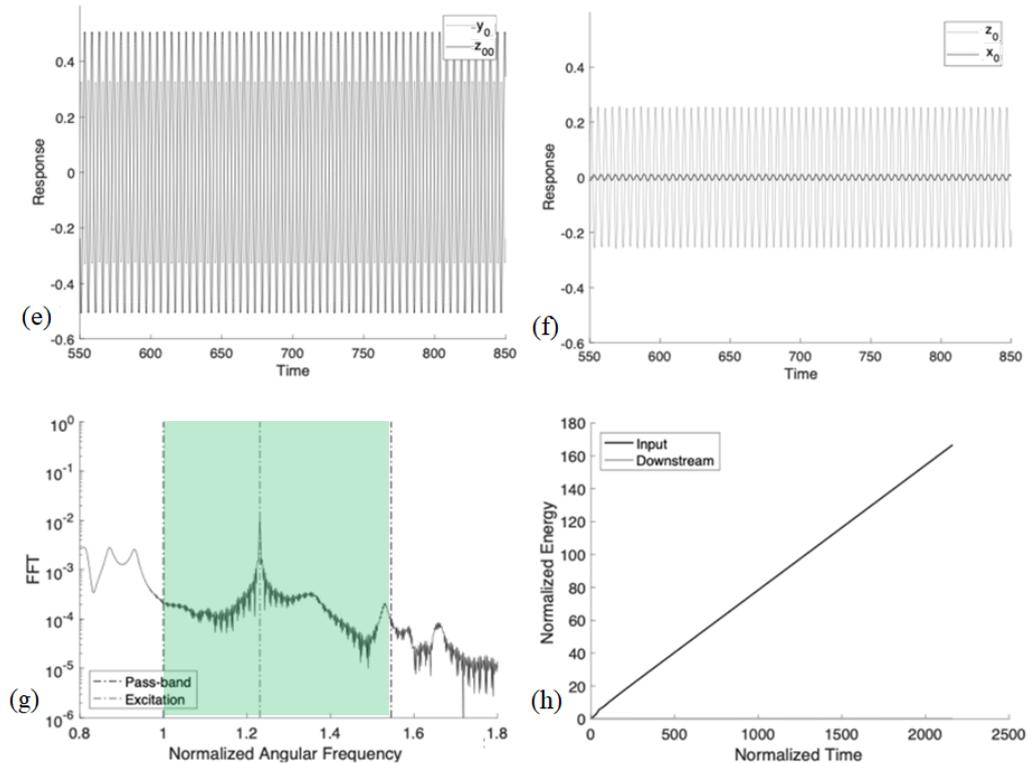

Figure 3. L-R wave propagation for System 2: (a,b) Time series of the oscillators of the nonlinear gates $(x_0,z_0)$ and $(z_{00},y_0)$, respectively, (c) Fourier spectrum of the oscillator $y_0$, and (d) energy measures; R-L wave transmission for System 2: (e,f) Time series of the oscillators of the nonlinear gates $(y_0,z_{00})$ and $(z_0,x_0)$, respectively, (g) Fourier spectrum of the oscillator $y_0$, and (h) energy measures – the pass band and excitation frequency are indicated in Figs. (c) and (g).

Lastly, in Figure 4 we depict the corresponding simulations for System 3. Again, energy transmission is only allowed only in the L-R direction, so the acoustics is non-reciprocal. This case reveals new physics in the gated waveguide under investigation, since for L-R propagation the responses appear to be highly irregular (chaotic), whereas for R-L propagation the responses are nearly monochromatic. To highlight in more detail the non-reciprocal acoustics, we define the following transmissibility measure, $\eta(T)$, and non-reciprocity measure, $\delta(T)$ [17],

$$\eta(T) = \begin{cases} \dfrac{E_{downL-R}(T)}{E_{input}(T)} & (L-R) \\ \dfrac{E_{downR-L}(T)}{E_{input}(T)} & (R-L) \end{cases} \quad (4)$$

$$\delta(T) = \log_{10} \dfrac{W_{downL-R}(T)}{W_{downR-L}(T)}$$

where the energy measures were defined in (3), and $T = 1,500$ denotes the total normalized time of the simulation. Note that reciprocity corresponds to $\delta = 0$ [28], so $\delta$ measures the difference (in logarithmic scale) in the transmitted energy downstream when the excitation is switched between the L and R sub-waveguides. In addition to maximizing acoustic non-reciprocity, we are also interested in maximizing the transmissibility measure $\eta$, which is defined as the ratio of the energy transmitted downstream to input energy. Lastly, we note that since the measures $\eta$ and $\delta$ converge at steady state, the normalized time interval $T$ is selected sufficiently large such that the acoustics reach steady state.

Returning to the non-reciprocal responses of System 3, we compute the non-reciprocity measure as $\delta = 8.435$, and the transmissibility measure $\eta = 0.26$. Considering the achieved non-reciprocity and comparing it to the previously reported non-reciprocity measure for the same waveguide but incorporating a single nonlinear gate [17], it is a marked improvement. Specifically, the maximum non-reciprocity measure achieved in [17] was $\eta \sim 3.0$, which is orders of magnitude less than the current result considering that the measure is in logarithmic scale; moreover, it is worth noting that this highly intense non-reciprocity is achieved while maintaining similar levels of transmissibility to [17].

The displacement time series are examined in detail in Figure 4 for the strongly non-reciprocal responses observed in System 3. From Figs. 4a and 4b, we note that the oscillation amplitudes of $x_0$ and $y_0$ are comparable for L-R propagation. However, when the excitation is switched to the opposite side (cf. Figs. 4e,f), the oscillation amplitude of $z_0$ significantly decreases so without loss of generality, we may introduce the scalings $x_0 \sim O(1)$ and $z_0 \sim O(\varepsilon)$, where $0 < \varepsilon \ll 1$ is a small perturbation parameter. Since the coupling stiffnesses at the nonlinear gates are essential nonlinear (i.e., they possess with zero – or nearly zero – linearized stiffness), the interaction forces at the gate $(z_{00}, y_0)$ are of $O(\varepsilon^3)$, and, therefore, $y_0 \sim O(\varepsilon^3)$. Such drastic reduction in the amplitudes of the gate oscillators that yields drastically low transmissibility and high non-reciprocity is only realizable in the waveguide with multiple (in this particular work, two) essentially nonlinear (i.e., non-linearizable) gates.

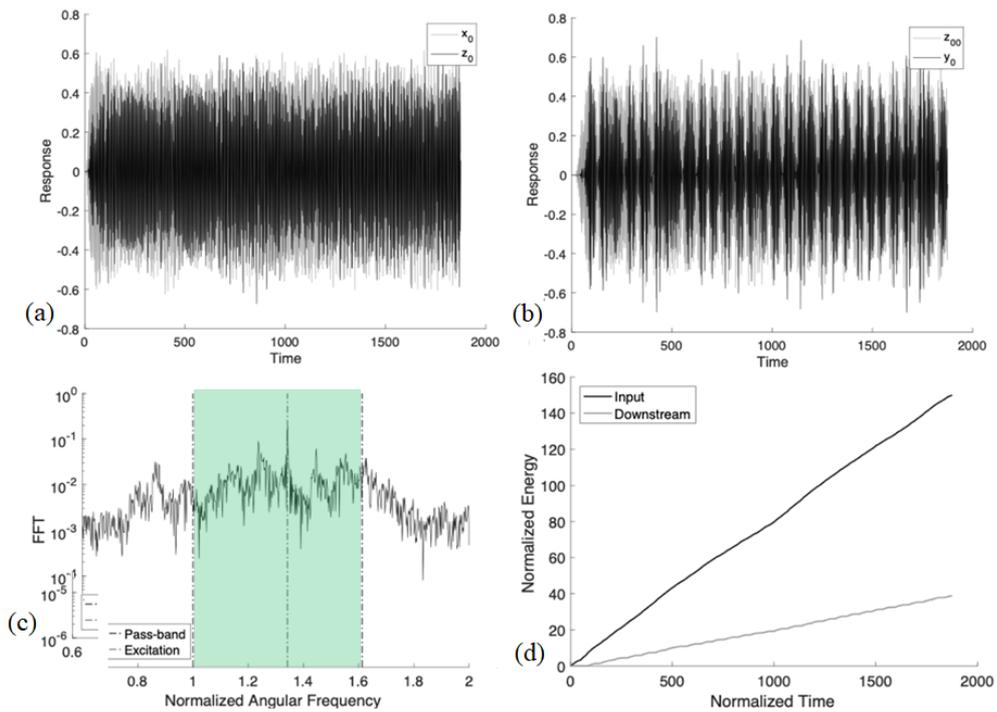
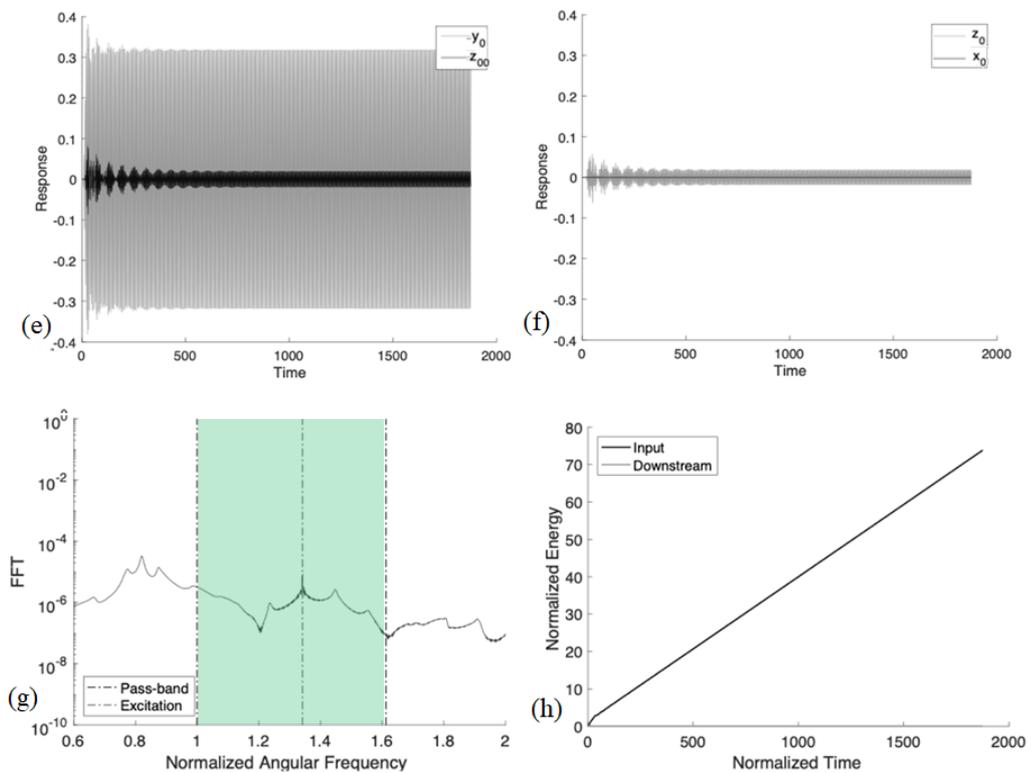

Figure 4. L-R wave propagation for System 3: (a,b) Time series of the oscillators of the nonlinear gates $(x_0, z_0)$ and $(z_{00}, y_0)$, respectively, (c) Fourier spectrum of the oscillator $y_0$, and (d) energy measures; R-L wave transmission for System 2: (e,f) Time series of the oscillators of the nonlinear gates $(y_0, z_{00})$ and $(z_0, x_0)$,, respectively, (g) Fourier spectrum of the oscillator $y_0$, and (h) energy measures – the pass band and excitation frequency are indicated in Figs. (c) and (g).

Motivated by these preliminary numerical results, we continue to investigate further the existence of acoustics with extremely strong non-reciprocity. This was performed by performing numerical simulations of the gated waveguide (1) with varying system parameters in the desired parameter space. Approximately 60,000 simulations we performed of which 35% were randomly chosen to populate the diagram of Figure 5, where the resulting non-reciprocity measure $\delta$ is plotted against the transmissibility measure $\eta$. In these simulations the nonlinear parameters are in the range $(\alpha_1, \alpha_2) \in [0,6]$, the coupling stiffness in the range $d \in [0.2,0.6]$, the forcing magnitude in the range $A_p \in [0.2,1]$, and the wavenumber in the range $\theta \in [0, \pi]$. The dataset is populated with parameters within these ranges, which are generated by randomly sampling from a uniform distribution. The rest of the parameters are kept constant. The result clearly shows three distinct branches of solutions with respect to the corresponding non-reciprocity measure. Focusing on the region with transmissibility measure, $\eta > 0.2$, we define the three branches of solutions, namely, (i) a highly non-reciprocal branch (HNRB) corresponding to $5 < \delta < 9$, (ii) an intermediate non-reciprocal branch (INRB) for $1 < \delta < 5$, and (iii) a near-reciprocal branch (NRB) for $[\delta] \ll 1$ (cf. Fig. 5). It should be clear that these branches correspond to distinct nonlinear acoustics of the gated waveguide, so we proceed to study in more detail the non-reciprocal responses of the gated waveguide oscillators in an effort to highlight the role that the two gates play in the generation of the generated acoustic non-reciprocity.

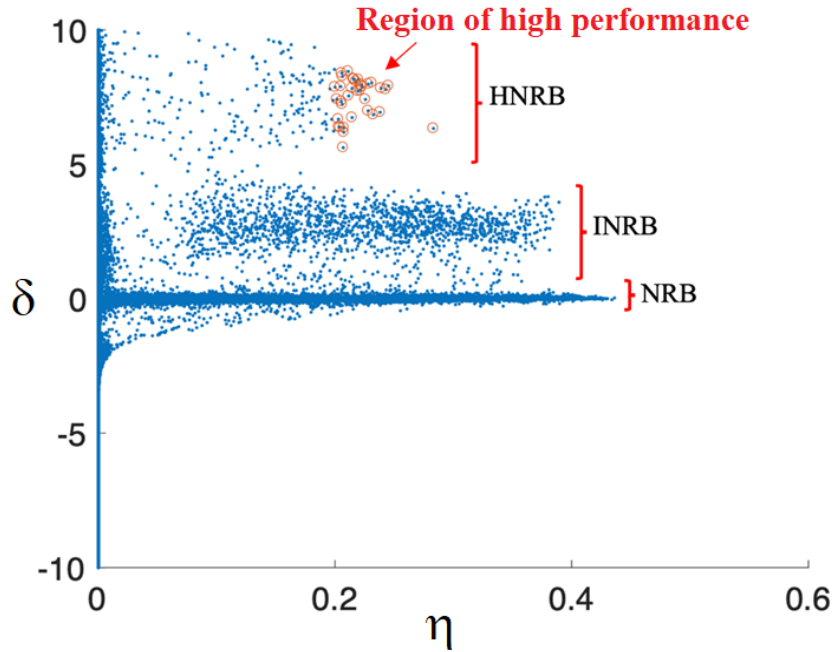

Figure 5. Non-reciprocity versus transmissibility diagram based on numerical simulations of the gated waveguide (1).

We initiate this study by comparing two cases corresponding to the same transmissibility from the branches characterized as HNRB (high non-reciprocity) and INRB (intermediate non-reciprocity), with the aim of gaining a better understanding of the qualitative differences in the acoustics yielding drastically different non-reciprocity measures. We note that the acoustics of System 3 (considered in Fig. 4) is in the HNRB branch, so it constitutes a first data set in our comparative study. For a case in branch INRB we consider the waveguide with system and forcing parameters $(\alpha_1, \alpha_2) = [1.7033, 3.0437], \sigma = -1.4, d = 0.275, \zeta = 0.013, A_p = 0.54$, and $\theta =$ $1.67\pi/6$, corresponding to $\delta = 2,42$, and $\eta = 0.26$ (i.e., identical to the transmissibility measure of System 3). We designate this case as System 4.

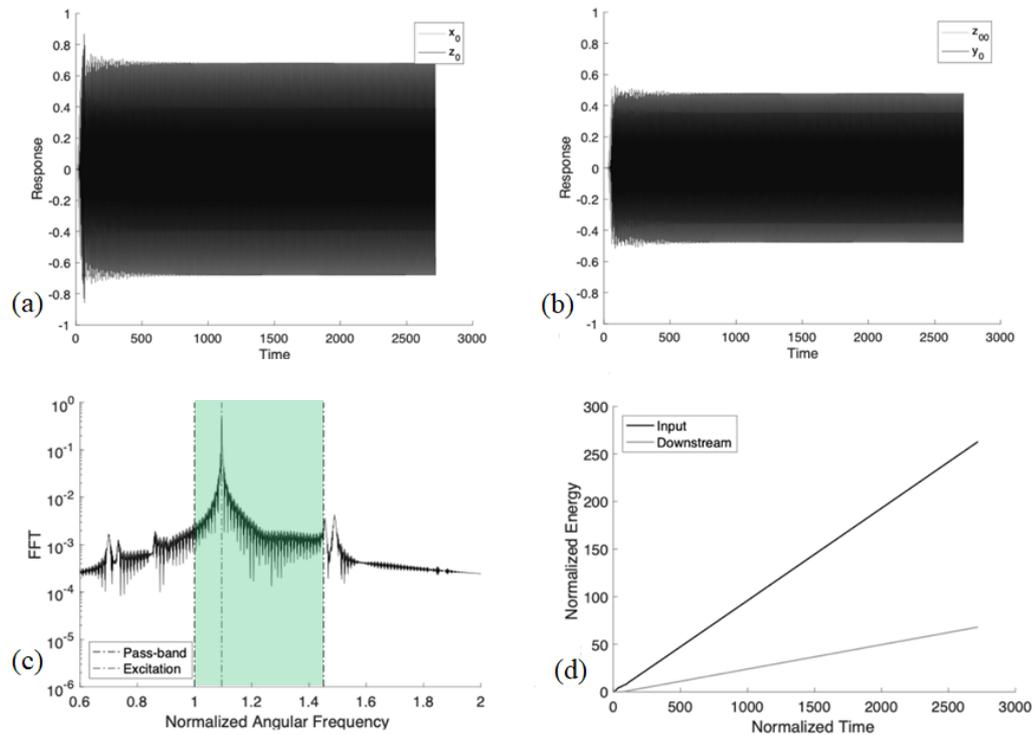

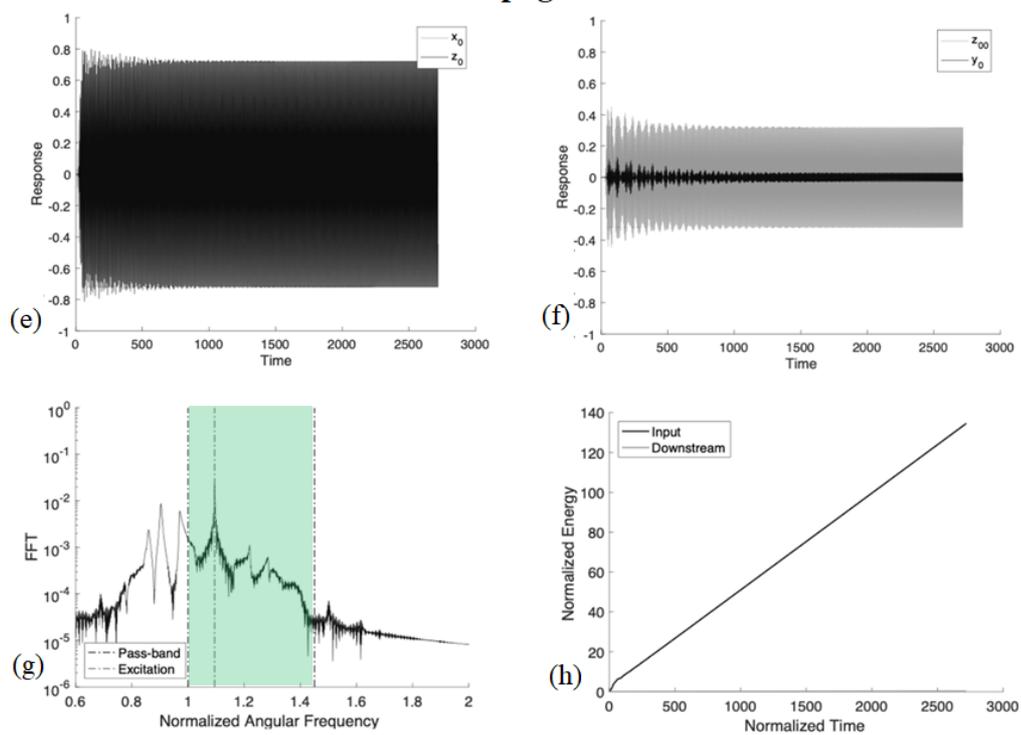

Figure 6. The nonlinear acoustics of System 4 in branch INRB: (a-d) L-R propagation, and (e-h) R-L propagation (caption as in Figs. 2-4).

A first remark is that the responses of System 4 depicted in Fig. 6 are dominated by a single harmonic, which is identical to the excitation frequency; hence, the acoustics is nearly monochromatic. Moreover, comparing the responses of Systems 3 and 4, we proceed to identify the different governing nonlinear acoustics of the waveguide responses in the HNRB and INRB branches. The qualitative difference between these cases can be identified by computing the total energies in (i.e., kinetic and potential energies of the component oscillators of) the L and R sub-waveguides and the central region (cf. Fig. 1), since this will reveal the flow of energy from the upstream (i.e., the sub-waveguide where the harmonic excitation is applied) to the downstream through the two nonlinear gates. In Figure 7, we depict the instantaneous energies of the sub-waveguides and the central region for Systems 3 and 4, for both L-R and R-L wave propagation. Notice, that in both cases there is more energy transmitted downstream for L-R propagation compared to R-L propagation; this is due to the combined nonlinearity and asymmetry of the central region between the two gates. The main difference is that while in both Systems 3 and 4 the energy that transmits downstream is low for R-L propagation, the HNRB system has almost all the energy concentrated to the upstream sub-waveguide; this occurs because of the reflection back to the upstream of the impeding energy at the central region between the two nonlinear gates. This nearly complete reflection of energy at the gates appears to be the main cause of the high non-reciprocity measure in System 3 compared to System 4.

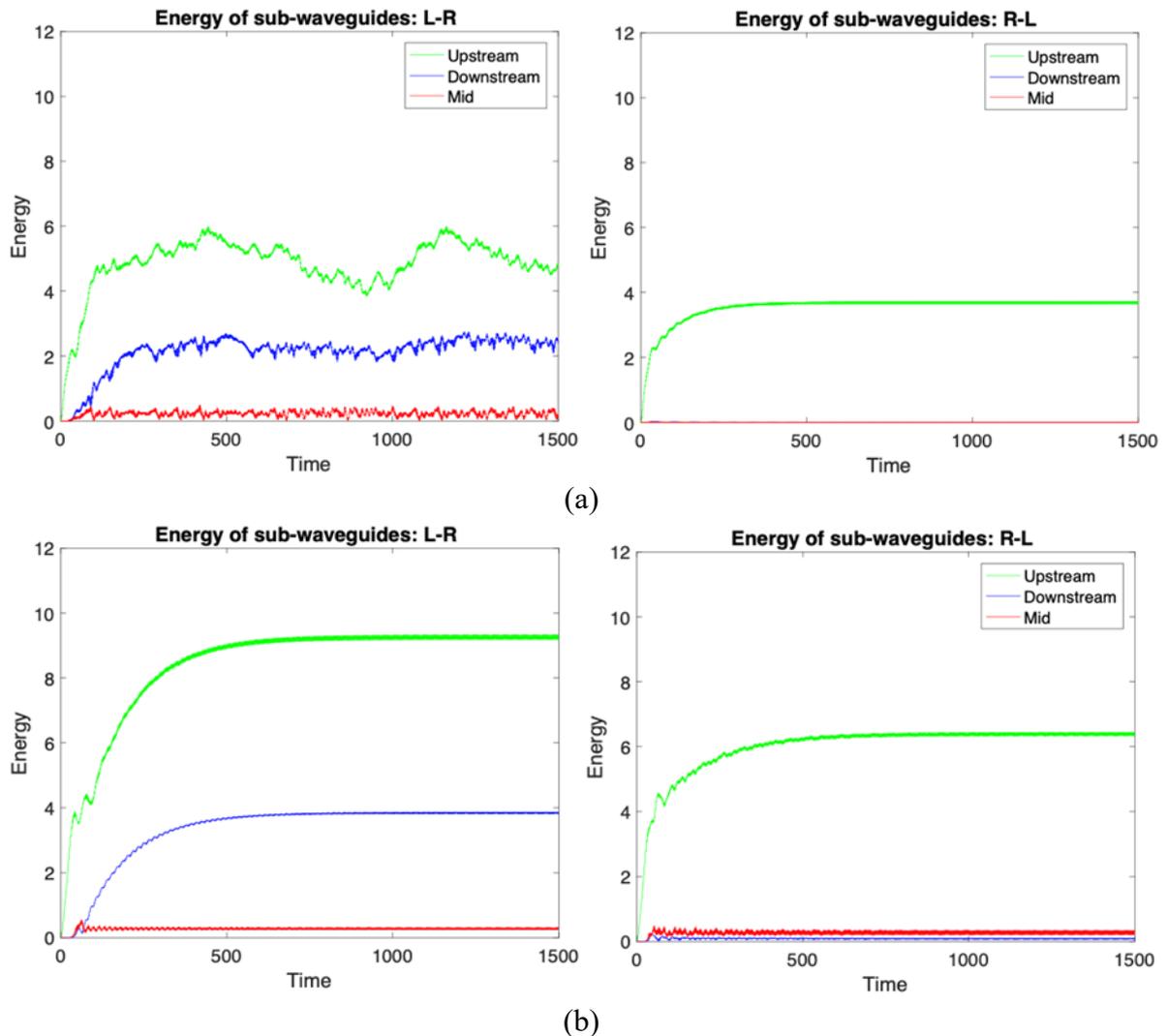

Figure 7. Instantaneous energies in the L and R sub-waveguides and in the central region between the two gates for (a) System 3 and (b) System 4; both L-R and R-L propagation are considered.

This qualitative difference in the acoustics can be observed also by comparing the responses of Systems 3 and 4 in Figs. 4 and 6, respectively. Indeed, in Figs. 4e,f, we note that the oscillations of the gate oscillators $z_0$ and $z_{00}$, have much smaller amplitudes compared to the corresponding ones in Figs. 6e,f. This comparison supports the conclusion, that in the HNRB case a larger amount of energy gets "trapped" in the upstream sub-waveguide compared to the INRB case. In System 3, most of the input energy reflects at the first nonlinear gate (in the order encountered by the impeding waves from the upstream), whereas in System 4 that occurs at the second nonlinear gate.

The previous results provide ample motivation to explore further and in more detail the non-reciprocal acoustics of the two-gated waveguide, especially given that the intensity of the achieved non-reciprocity can be orders of magnitude higher compared to the one in the corresponding waveguide with a single gate [17]. To this end, in the next section we will resort to machine learning to account for the high dimensionality of the parameter space of the problem. However, as a last step in our preliminary simulations, we explore in more detail the effect of the damping at the gate oscillators on the acoustics. In Figure 8 we depict the non-reciprocal transmissibility $\eta$ and downstream energy $E_{down}$ measures when the normalized damping ratio $\zeta$ varies from 0 to 0.05 for Systems 1, 2 and 3, and for both L-R and R-L wave transmission. These results show clearly the important effect that viscous dissipation of the gates has on the acoustics.

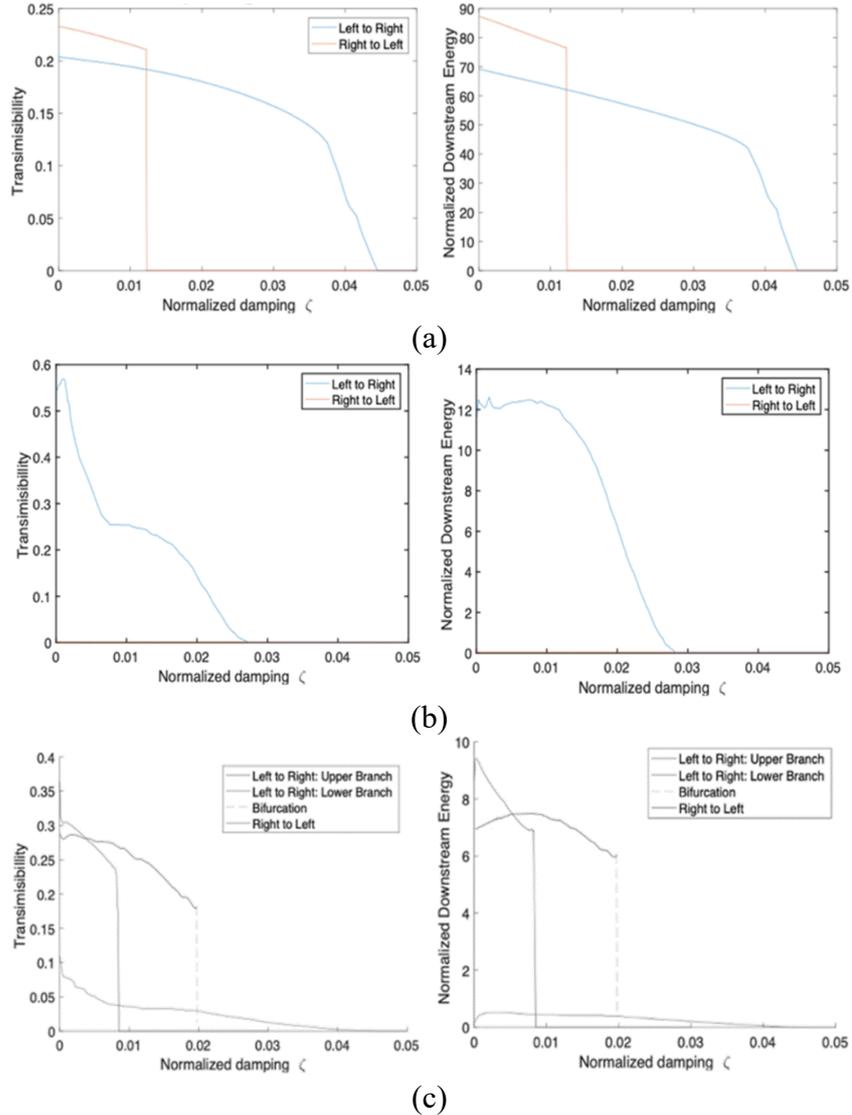

Figure 8. Transmissibility and downstream energy measures in the double gated waveguide for varying normalized damping ratio $\zeta$: (a) System 1, (b) System 2 and (c) System 3.

Considering the results for System 1 (cf. Fig. 8a) the transmissibility and downstream energy measures decrease when the normalized damping ratio $\zeta$ increases, and, in addition, bifurcations are observed at critical values of damping. This is similar behavior to the case of the one gated waveguide [17], where steep descents in downstream energy are observed for both L-R propagation and R-L propagation at certain critical thresholds of the damping ratio. The critical damping ratios for the bifurcations are $\zeta_{cr\_RL} \sim 0.012$ for R-L propagation and $\zeta_{cr\_LR} \sim 0.04$ for L-R propagation. As noted in [17], this disparity in the critical damping ratios provides a region of strong non-reciprocity, where energy can be transmitted from left to right but is prohibited from propagating right to left. This is another advantage that we can note for the system with the two nonlinear gates compared to the one presented in [17] with a single gate, since we see that there is a larger design space with respect to the normalized parameter $\zeta$ were we can take advantage of the realization of high acoustic non-reciprocity.

Another interesting acoustic feature is observed from the responses of System 2 (cf. Fig. 8b). Indeed, we observe two steep descents in the transmissibility plot for L-R propagation at two critical damping ratios; at the same time, there negligible transmissibility for R-L propagation, providing a large range (in terms of the damping ratio) for the realization of strong non-reciprocity in the acoustics. These results provide different design options with respect to transmissibility and non-reciprocity for this gated waveguide configuration.

Lastly, considering the results for System 3 (cf. Fig. 8c) which corresponds to cases in the HNRB branch we note further interesting features in the nonlinear acoustics. Besides the already familiar bifurcations occurring with varying normalized damping ratio $\zeta$, in this case there exist two co-existing branches solutions for L-R propagation (but only a single branch for R-L propagation). These co-existing solutions are generated solely by the strong nonlinearity at the gates, and which one is realized depends on the initial conditions. Clearly, the existence of co-existing solutions as well as the very high acoustic non-reciprocity achieved in this system (which is orders of magnitude higher compared to what is possible for the case of a single gate [17]), provides additional opportunities for *robust design of the two-gated waveguide for simultaneous high transmissibility and high acoustic non-reciprocity*. At this point it is important to highlight the exciting prospects that the extremely high non-reciprocal branch of solutions opens. In our work, the harmonic excitation is applied at an intermediate point of the lattice (not at the end of the lattice), so without the nonlinear gate, the lattice is homogeneous and 50% of the input energy propagates to the upstream and 50% the downstream. Therefore, the theoretical limit of the transmissibility is $\eta_{max} = 0.5\ \eta$. Using the specific combination of nonlinear gates of System 3 we achieve an extremely high measure of non-reciprocity, $\delta \sim 10$, while at the same time maintaining a high transmissibility measure 0.2 in all cases shown in Fig. 8c.

To the authors best knowledge, there is no reported active or passive, linear or nonlinear non-reciprocal device in the current literature that is capable of achieving these levels of performance [18, 19, 20] . In previous work [15, 17], it was proven how effective a single nonlinear gate could be for the waveguide under consideration. In this work, we show that using a combination of two nonlinear gates, we can increase by three times the non-reciprocity measure, which translates to an energy discrepancy in the ratio of $10^9$ based on (4). This result indicates that the use of multiple local nonlinear gates in an extended linear waveguide can yield extremely intense acoustic non-reciprocal acoustics, without sacrificing the levels of transmissibility in the desired (preferred) direction of wave propagation. In a broader perspective, the current two-gated waveguide may be regarded as a case in a broader class of systems incorporating cascades of hierarchical nonlinear local gates, and the methods developed in this work (as well as the analytical methods discussed in [15, 17]) can be extended for this broader class. Moreover, the insight in the nonlinear acoustics presented herein and the machine learning approach discussed in the next section could be applied to the design of a broad class of nonlinear phononic lattices subject to harmonic or transient (e.g., shock) excitation.

## 3. Machine Learning Approach for Predictable Design

In this section a learnable simulator is built to predict reliably the transmissibility and non-reciprocity of the two-gated waveguide in a multi-dimensional parameter space. In addition to the normalized excitation magnitude, $A_p$, we consider four additional normalized independent parameters for the nonlinear acoustics, namely, the nonlinear coefficients of the gate oscillators, $\alpha_1, \alpha_2$, the excitation frequency $\hat{\omega}$, and, lastly, the (strong) intra-waveguide coupling coefficient, $d$. In the following exposition, the detuning (asymmetry) coefficient is kept fixed to $\sigma = -1.4$, whereas, the damping ratio of the gate oscillators varies with the normalized coupling stiffness $d$, that is, $\xi = d\zeta$, with $\zeta = 0.013$. Note that due to the strongly nonlinear gates that acoustics is expected to be energy (or excitation) dependent, so including $A_p$ as one of the design variables is important since different acoustic phenomena may be realized at different excitation levels. Moreover, following the assumptions made in previous works regarding non-uniform damping distribution in the linear waveguide [14, 15, 17], we assume that the viscous damping ratios of the grounding stiffnesses of the component linear oscillators are uniformly equal to $\zeta = 0.013$. Employing machine learning we aim to predictively identify the values of these parameters for optimal (maximum) non-reciprocity and transmissibility of the waveguide in the five-dimensional parameter space $(A_p, \alpha_1, \alpha_2, \hat{\omega}, d)$.

To apply the machine learning approach, a dataset is created with five variables. The aim is to restrict the excitation frequency $\hat{\omega}$ within the range of the passband of the linear (ungated) waveguide defined by the limiting values of the wavenumber $\theta = 0, \pi$. We note, however, that if $\hat{\omega}$ is selected close to the boundary of the passband, the corresponding group velocity of propagating waves becomes very small, and it takes a long time for the acoustics to reach a steady state. In addition, the nonlinear parameters are restricted in the range $(\alpha_1, \alpha_2) \in [0, 6]$, and the other two parameters in the range $A_p \in [0.2, 1]$ and $d \in [0.2, 0.6]$. The dataset contains approximately 60,000 sets of these parameters, which are generated by randomly sampling from a uniform distribution.

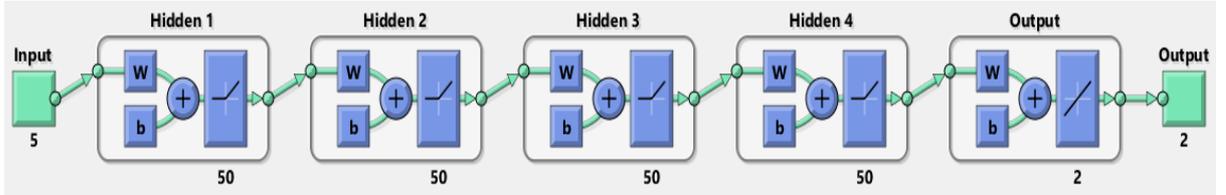

Figure 9. Schematic of the NN model.

A machine learning model is developed as a cost-effective alternative to direct numerical simulations of a strongly coupled gated waveguide (1). The model is based on the previous similar models developed in [15, 17] and is capable of significantly reducing the required simulation time while, at the same time, maintaining acceptable accuracy and reliability of the predicted nonlinear acoustics. The model takes as inputs the five variables $(A_p, \alpha_1, \alpha_2, \hat{\omega}, d)$, and provides as outputs the non-reciprocity and transmissibility measures $\delta$ and $\eta$, respectively. Initially, the machine learning model is trained using an artificial neural network (NN) and the above-mentioned dataset. The inputs and outputs are normalized and assembled into vectors of dimensions (5 x 1) and (2 x 1), respectively, with the minimum value mapped to $-1$ and the maximum to $+1$. The NN is composed of 4 hidden layers each with 50 neurons (cf. Figure 9), which as in [17] employs the ReLU activation function for the output layer. Specifically, the normalized input is mapped to the normalized output as shown in Fig. 9, and the estimates of each layer are computed iteratively from the previous layer as, $\boldsymbol{a}_i = f(\boldsymbol{w}_i \boldsymbol{a}_{i-1} + \boldsymbol{b}_i)$, where $\boldsymbol{a}_i$ denotes the vector of the values at the $i-th$ layer, $\boldsymbol{w}_i$ and $\boldsymbol{b}_i$ are the learnable weight matrix and the offset vector, respectively, and $f(\cdot$

) is the activation function. As in [17] the activation functions of the hidden layers are taken as rectified linear unit (ReLU) functions, where $f(x) = \max(0, x)$, and the activation function of the output layer is linear (identity transformation). The overall NN model is developed and trained using the MATLAB® deep learning toolbox.

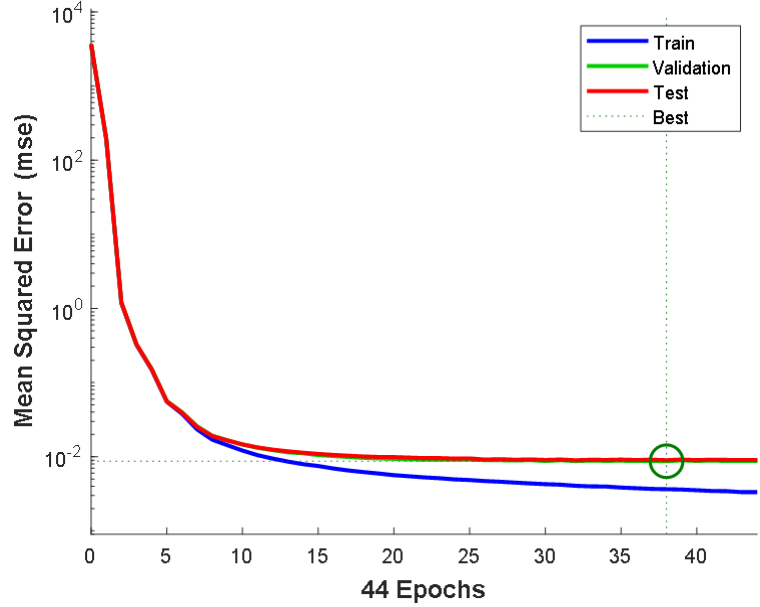

Figure 10. Convergence study: Loss function of the NN at each epoch.

The overall dataset is randomly divided into three sub-sets with 50% used for training, 15% for validation and 35% for testing. The Levenberg-Marquardt backpropagation algorithm is adopted to train the weights $\boldsymbol{w}_i$, and the offsets $\boldsymbol{b}_i$, with the mean square error adopted as the loss function. The loss function for the validation sub-set is evaluated after each epoch (defined as the number of passes of the entire training set), and the training stops if the loss function on the validation subset does not improve after six consecutive epochs. The loss function for each epoch is shown in Figure 10, and this convergence study indicates that the minimum loss function on the validation set (equaling 0.0087) was detected at epoch 38, indicating that the NN is properly trained. Following similar validation steps as discussed in [17] the effectiveness of the trained NN simulator is evaluated by plotting the predicted non-reciprocity and transmissibility measures, and comparing the results with those from direct numerical simulations on the testing subset; this shown in Figure 11, where the cases in the desired region corresponding to $0.2 < \eta$ and $5 < \delta$ are highlighted in the scatter plot. From these results it is clear that the NN accurately predicts the non-reciprocity and transmissibility of the gated waveguide on the training set, since the correlation values $R$ between the targets generated by the simulations and the outputs predicted by the NN simulator are greater than 0.96. Indeed, from Fig. 11d we note that the three distinct branches of high non-reciprocity (HNRB), intermediate non-reciprocity (INRB) and near-reciprocity (NRB) are fully recovered on the testing set. Hence, the NN simulator accurately captures the nonlinear acoustics of the gated waveguide, but at much less computational expense compared to direct numerical simulations, so its efficacy is established. Hence, the NN model can now be employed to optimize the nonlinear acoustics of the gated waveguide in the designated 5-dim parameter space in terms of maximum non-reciprocity with desirable transmissibility.

At this point we note that our aim is to conceive a non-reciprocal gated waveguide with very high values of both non-reciprocity and transmissibility measures in the preferred direction where energy is transmitted. In particular, it is desirable to have $\eta > 0.2$ and $\delta$ as large as possible, meaning that at least 20% of the input energy should be transmitted downstream in the preferred direction of wave propagation, and that the transmitted energy in that direction should be at orders of magnitude larger than in the opposite direction. Indeed, from the results depicted in Figs 5 and 11, it is achievable to obtain high values for $\eta$ and $\delta$ in the ranges $0.2 < \eta < 0.4$ and $5 < \delta < 10$, i.e., in the desired branches HNRB of Figs. 5 and 11.

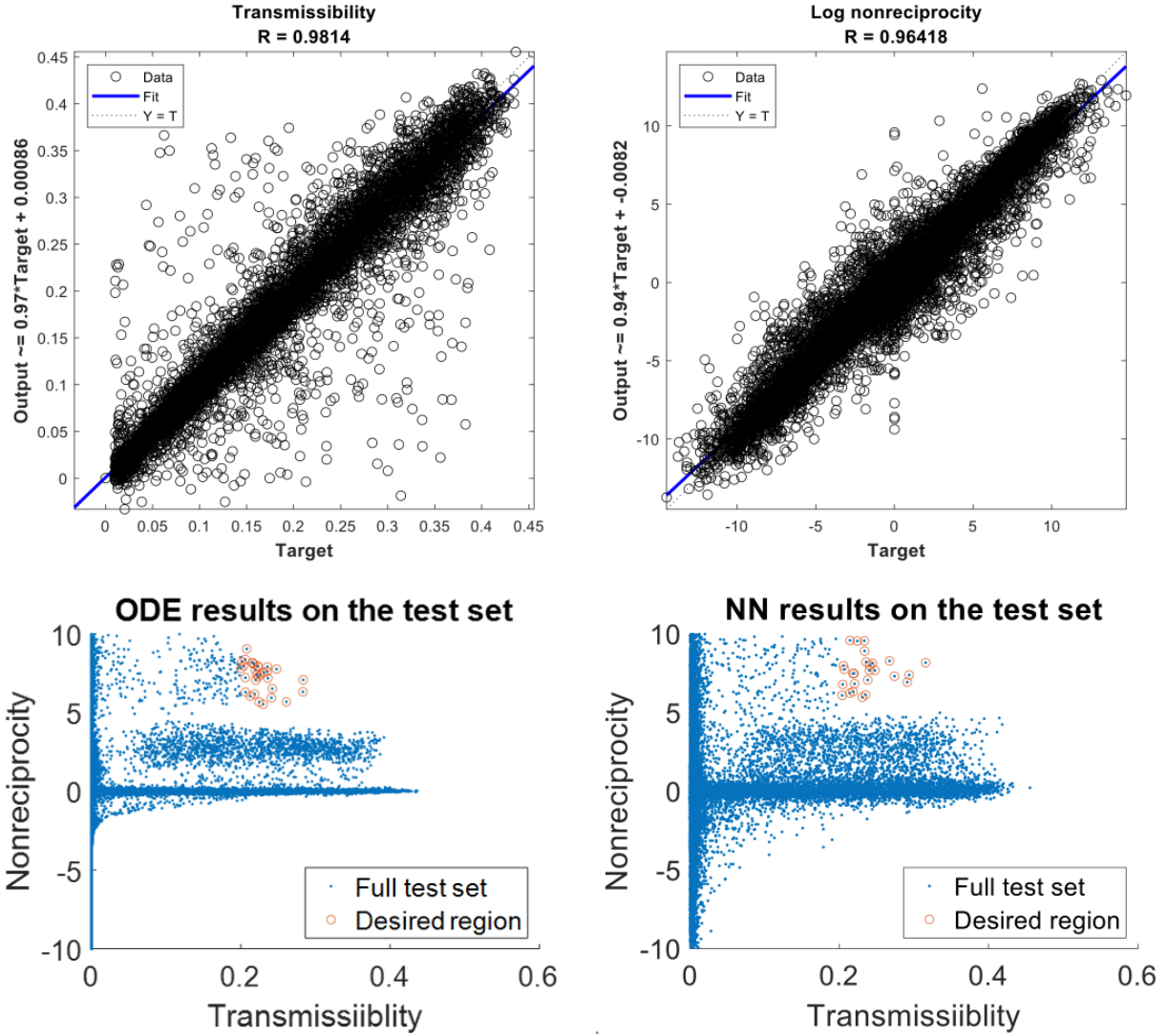

Figure 11. Correlation between the target and output (prediction) of the test set: (a) transmissibility measure, (b) non-reciprocity measure, and non-reciprocity versus transmissibility plots based on (c) direct numerical simulations of (1) and (d) NN predictions.

Based on the outputs of the trained NN simulator on the test set, we employ the same metrics as in [17] to fill the confusion matrix shown in Table 2. As in [17] it classifies the test subset into four distinct groups, namely, true positive (TP), true negative (TN), false positive (FP) and false negative (FN). Specifically, TP denotes the amount of true positive values that are desirable both in terms of the NN simulator and the direct numerical simulations; TN denotes the amount of true negative values that are undesirable by both the NN simulator and the numerical simulation; FP is the amount of false positive values that are found to be desirable by the NN simulator but undesirable by the numerical simulations; and FN denotes the amount of false negative values that are undesirable by the NN simulator but desirable by the numerical simulations. Based on these designations, the sensitivity, specificity and precision measures are then defined as follows:

$$\text{Sensitivity} = \frac{TP}{TP + FN}, \quad \text{Specificity} = \frac{TN}{TN + FP}, \quad \text{Precision} = \frac{TP}{TP + FP} \quad (5)$$

From Table 2 we note that the sensitivity, precision, and specificity measures are all close to unity, so we conclude that the NN simulator is reliable, in the sense that it predicts well the class of two-gated waveguides exhibiting large transmissibility and non- reciprocity measures.

Table 1. Confusion matrix for the predictions of the NN on the test subset.

| Data points in the test set: $N = 21,000$ | Predicted condition | | Performance |
|---|---|---|---|
| Actual condition | Desirable<br>TP = 726<br><br>FP = 165 | Undesirable<br>FN = 160<br><br>TN = 19,949 | Sensitivity:<br>0.8254<br>Specificity:<br>0.9920<br>Precision:<br>0.8160 |

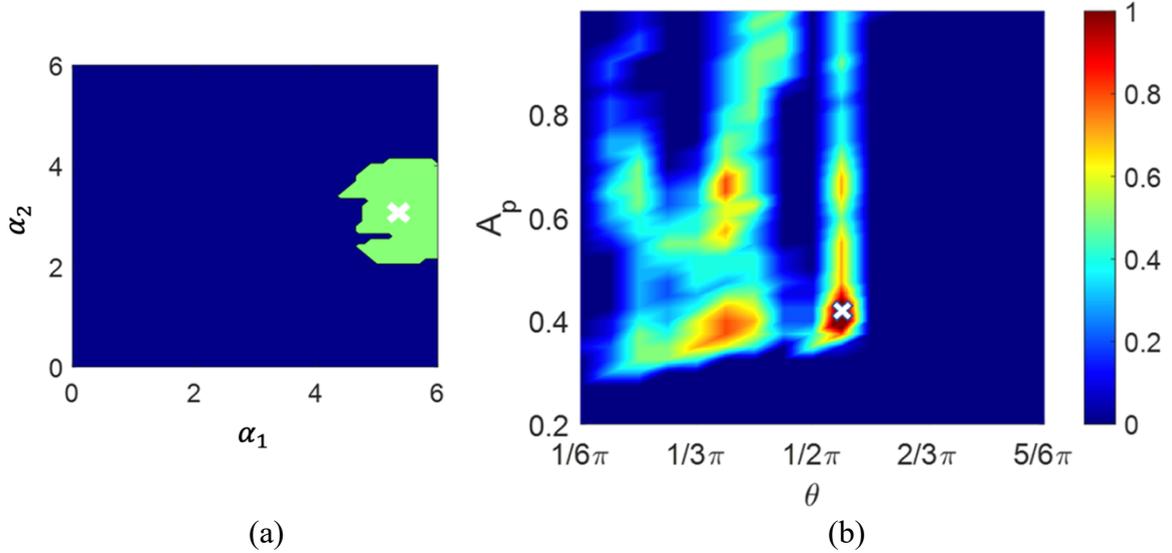

(a)          (b)

Figure 12. NN simulator predictions: (a) Desired parameter region in the $(\alpha_1, \alpha_2)$ plane for $(d, A_p, \theta) = (0.2, 0.4, 3.2659/6\pi)$, and (b) contour plot of kernel size at the desired region in $(\alpha_1, \alpha_2)$ for $d = 0.2$; the contour plot of (b) was constructed only taking into account the cases corresponding to $\delta > 4$ and $\eta > 0.2$.

To highlight the results obtained by the NN simulator, in Figure 12 we provide a visualization of the predicted performance classification based on the NN output on the test set. The case considered in Fig. 12a corresponds to $(d, A_p, \theta) = (0.2, 0.4, 3.2659/6\pi)$ and varying nonlinear coefficients $(\alpha_1, \alpha_2)$. The predicted region of desired performance is shown in light green, where we select the case denoted by (×) corresponding to $\alpha_1 = 5.0$ and $\alpha_2 = 3.1$. To visualize the desired region in higher parameter space, a square is fit within it with its side length labelled as the "kernel size"; this is then employed to quantify the robustness of the desired performance when additional parameters vary. Indeed, in Fig. 12b, the normalized intra-waveguide coupling stiffness is fixed to $d = 0.2$, and a contour plot of the kernel size as a function of $A_p$ and $\theta$ is depicted. For these values we proceed to construct the contour plot of the kernel size depicted in Fig. 12b for $d = 0.2$, and select a case with optimal robustness for $A_p = 0.4$ and $\theta = 3.2659/6\pi$ corresponding to the point indicated by (×).

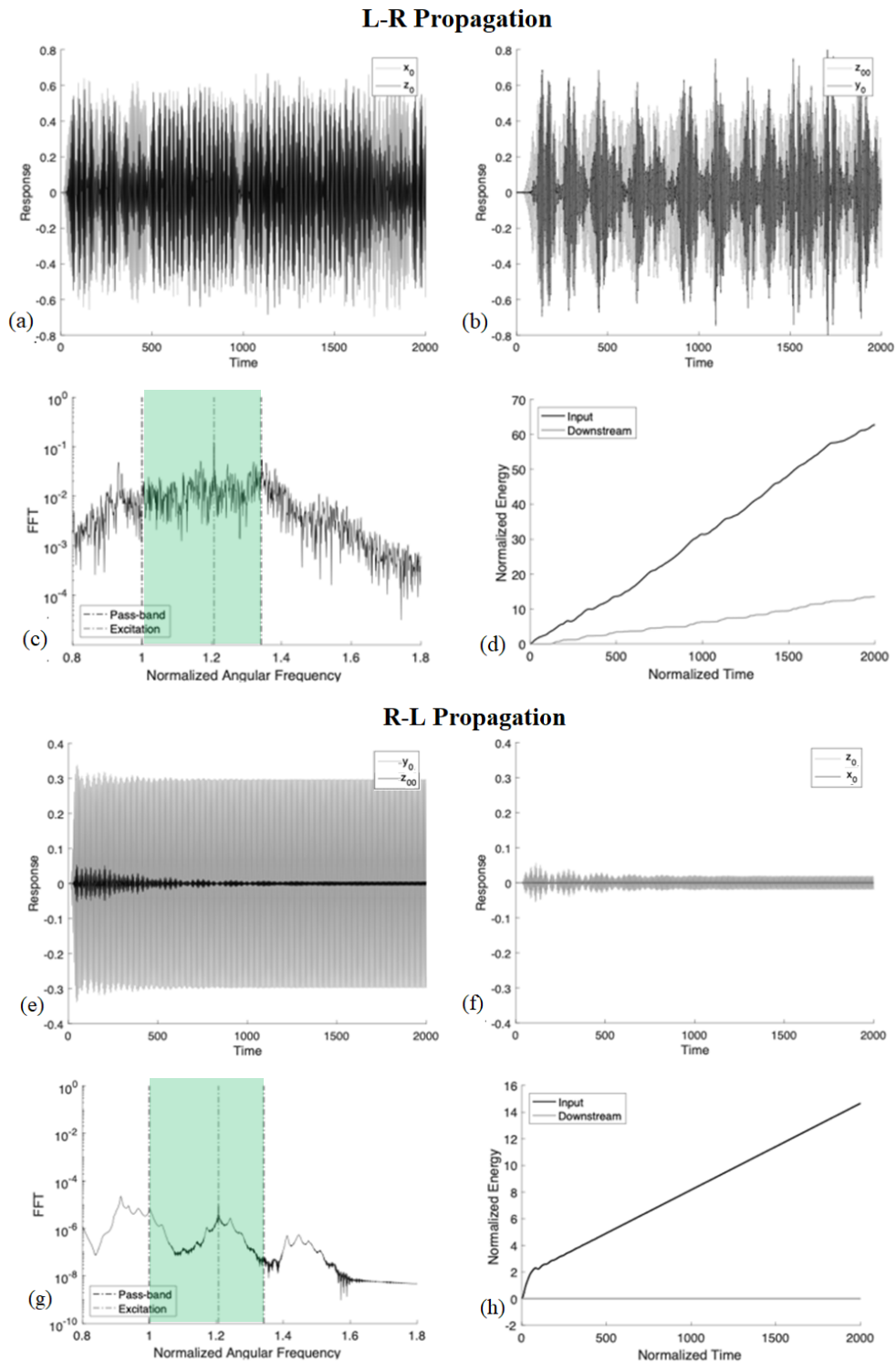

Figure 13. The nonlinear acoustics of the case corresponding to (×) in Fig. 12: (a-d) L-R propagation, and (e-h) R-L propagation (caption as in Figs. 2-4).

The predicted measures of transmissibility and non-reciprocity for this case are $\eta = 0.2257$ and $\delta = 8.2700$, respectively. In Figure 13 we depict the responses at the nonlinear gates for this case and infer the extremely strong non-reciprocity in the acoustics, and the good transmissibility achieved in the preferred L-R direction of propagation. This last example demonstrates that the machine learning approach presented in this work can provide a reliable predictive tool for designing acoustic waveguides for maximum non-reciprocity, achieving at the same time good transmissibility in the preferred direction of wave propagation. Such tools can be employed to conceive practical non-reciprocal waveguides and acoustic metamaterials that incorporate local strong nonlinearities and asymmetries. Depending on the needs of the application, NN simulators can be utilized to determine the appropriate system parameters (even in relatively high dimensional spaces) based on the desired levels of transmissibility, nonreciprocity and robustness of such waveguides.

## 4. Concluding Remarks

This work studies the generation of strong acoustic non-reciprocity in a 1D passive linear waveguide which incorporates two strongly nonlinear, asymmetric gates. Harmonic excitation is applied to a single oscillator of the waveguide. Strong stiffness coupling between the waveguide's oscillators is assumed, yielding a passband for the undated waveguide, where traveling waves can propagate unattenuated in both left-to-right (L-R) and right-to-left (R-L) directions. The *local* nonlinear asymmetric gates break the symmetry and linearity of the waveguide resulting in *global* strong non-reciprocal acoustics. Moreover, to the authors' best knowledge, the considered two-gated waveguide may yield extremely strong acoustic non-reciprocity, much stronger compared to what is reported by current active or passive devices; moreover, this extreme break of reciprocity combines with desirable transmissibility in the preferred direction of wave propagation. Perhaps even more interesting, the analysis of the nonlinear acoustics uncovered the rich physics governing the break of reciprocity, which, itself can be the steppingstone for further analysis.

We demonstrated through numerical analysis that acoustic non-reciprocity in the waveguide can be realized either by monochromatic responses or by strongly modulated responses. Furthermore, machine learning was used for predictive design in terms of transmissibility and non-reciprocity to reduce computational time for high-dimensional parameter space analysis. The neural net (NN) model proved to be capable of capturing the waveguide responses with respect to defined measures of non-reciprocity and transmissibility. In the desired regions of performance, the transmissibility measure reached as high as 40% (out of a maximum of 50% for the ungated waveguide), while the non-reciprocity measure reached as high as 9, results that surpass the performance of previous non-reciprocal active and passive devices [15, 17, 18, 19, 20]. Moreover, compared to a similar study performed for the same waveguide but with a single gate [17], the reported levels of non-reciprocity herein exceed the previous results of that work by as much as 9 orders of magnitude. Moreover, the NN simulator enabled a study of the robustness of the strongly non-reciprocal responses to changes in system and excitation parameters.

In synopsis, through this study we investigate new theoretical and practical aspects on the application of strong nonlinearity and asymmetry for passive break of reciprocity in an acoustic waveguide. It is shown that it is feasible to incorporate multiple local nonlinear gates in spatially extended linear waveguide to achieve high performance with respect to global break of reciprocity and transmissibility. Hence, the present study can be regarded as a specific case of a broader class of waveguides, whose acoustics can be tailored to non-reciprocity through the use of cascades of local hierarchical gates. A challenge in these problems is the high dimensionality of the design parameter spaces, but, as shown in this work, this can be addressed through tools of machine learning, which can, not only provide predictions of designs of high performance, but also can inform studies of robustness of such solutions. Of course, this creates the need for large training sets and insightful pre-sampling of the data based on analytical and numerical predictions that would help optimize the training of the NN simulations. Hence, a physics-informed machine learning approach seems to be an appropriate design tool for this type of complex problems.

# 5. References

[1] J. Achenbach, *Reciprocity in elastodynamics*, Cambridge, UK: Cambridge University Press, 2009.

[2] J.C. Maxwell, On the calculation of the equilibrium and stiffness of frames, *The London, Edinburgh, and Dublin Philosophical Magazine and Journal of Science*, 27, 294-299, 1864.

[3] K. Tsakmakidis, L. Shen, S. Schulz, X. Zheng, J. Upham, X. Deng, H. Altug, A.F. Vakakis and R.W. Boyd, "Breaking Lorentz reciprocity to overcome the time-bandwidth limit in physics and engineering," *Science,* 356:6344, 1260-1264, 2017.

[4] R. Fleury, D. Sounas, C. Sieck, M. Haberman and A. Alù, "Sound isolation and giant linear nonreciprocity in a compact acoustic circulator," *Science,* 343:6170, 516-519, 2014.

[5] S.A. Cummer, "Selecting the direction of sound transmission," *Science,* 343:6170, 495-496, 2014.

[6] S.A. Cummer, J. Christensen and A. Alù, "Controlling sound with acoustic metamaterials," *Nature Reviews Materials,* 1:3, 1-13, 2016.

[7] I. Grinberg, A.F. Vakakis and O.V. Gendelman, "Acoustic diode: Wave non-reciprocity in nonlinearly coupled waveguides," *Wave Motion,* 83, 49-66, 2018.

[8] A. Darabi, L. Fang, A. Mojahed, M. D. Fronk, A.F. Vakakis and M.J. Leamy, "Broadband passive nonlinear acoustic diode," *Physical Review B,* 99:21, 214305, 2019.

[9] G. Trainiti and M. Ruzzene, "Non-reciprocal elastic wave propagation in spatiotemporal periodic structures," *New Journal of Physics,,* 18:8, 083047, 2016.

[10] A.A. Maznev, A.G. Every and O.B. Wright, "Reciprocity in reflection and transmission: What is a 'phonon diode'?," *Wave Motion,* 50: 4, 776-784, 2013.

[11] B. Liang, X.S. Guo, J. Tu, D. Zhang and J.C. Cheng, "An acoustic rectifier," *Nature materials,* 9:12 , 989-992, 2010.

[12] C. Fu, B. Wang, T. Zhao and C.Q. Chen, "High efficiency and broadband acoustic diodes," *Applied Physics Letters,* 112:5, 051902, 2018.

[13] S.H. Mousavi, A.B. Khanikaev and Z. Wang, "Topologically protected elastic waves in phononic metamaterials," *Nature communications,* 6:1, 1-7, 2015.

[14] C. Wang, A. Kanj, A. Mojahed, S. Tawfick and A.F. Vakakis, "Experimental Landau-Zener Tunneling (LZT) for Wave Redirection in Nonlinear Waveguides," *Physical Review Applied,* 14:3, 034053, 2020.

[15] C. Wang, A. Mojahed, S. Tawfick and A.F. Vakakis, "Machine learning non-reciprocity of a linear waveguide with a local nonlinear, asymmetric case: weak coupling," *Journal of Sound and Vibration*, 537, 117211, 2022.

[16] A. Blanchard, T.P. Sapsis and A.F. Vakakis, "Non-reciprocity in nonlinear elastodynamics," *Journal of Sound and Vibration,* 412, 326-335, 2018.

[17] C. Wang, A. Mojahed, S. Tawfick and A.F. Vakakis, "Machine Learning Non-reciprocity of a Linear Waveguide with a Local Nonlinear, Asymmetric Gate: Case of Strong Coupling," *Journal of Computational and Nonlinear Dynamics*, in press.

[18] Y. Zhai, H. Kwon and B. Popa, "Active Willis metamaterials for ultracompact nonreciprocal linear acoustic devices," *Physical Review B,* 99, 220301(R), 2019.

[19] B. Popa and S.A. Cummer, "Non-reciprocal and highly nonlinear active acoustic metamaterials," *Nature Communications,* 5, 3398, 2014.

[20] N. Geib, A. Sasmal, Z. Wang, Y. Zhai, B. Popa and K. Grosh, "Tunable nonlocal purely active nonreciprocal acoustic media," *Physical Review B,* 103, 165427, 2021.

**Statements & Declarations**

**Funding**

The authors declare that no funds, grants, or other support were received during the preparation of this manuscript.

**Competing Interests**

The authors have no relevant financial or non-financial interests to disclose.

**Data Availability**

The datasets generated and/or analyzed during the current study are not publicly available, but are available from the corresponding author on reasonable request.